# *Enhanced Hydrogen Storage in Gold-doped Carbon Nanotubes: A first-principles study*

Shima Rezaie [a], David M. J. Smeulders, [a,b] Azahara Luna-Triguero,*[a,b]

[a] *Energy Technology, Department of Mechanical Engineering, Eindhoven University of Technology, P.O. Box 513, 5600 MB Eindhoven, The Netherlands.*

[b] *Eindhoven Institute for Renewable Energy Systems (EIRES), Eindhoven University of Technology, PO Box 513, Eindhoven 5600 MB, The Netherlands*

## Abstract

Sorbent materials are a promising alternative to advance hydrogen storage technologies. The general disadvantage is the relatively weak solid-gas interaction and adsorption energy, providing low gravimetric and volumetric capacities and extreme operational conditions. Here we propose Au-doped carbon nanotubes (CNTs) as an efficient alternative for reversible hydrogen capture at high temperatures. This work investigates the properties of several modified CNTs using density functional theory.

We analyze the binding and formation energies of the uniformed Au-doped CNTs and assess their adsorption capability. The hydrogen storage mechanisms of the nanostructures are studied in depth using partial density of states and charge transfer analysis showing that the increase of diameter has a positive effect on the outcome. Our findings show that the modified structures are able to capture from six to nine hydrogen molecules per gold atom, achieving volumetric capacities ranging from 154 to 330 g/l, surpassing the DOE target. In addition, the calculated desorption temperatures indicate high performance of Au-doped CNTs, obtaining hydrogen capture-release working conditions above 200 K.

## Introduction

Global energy demand is growing much faster than energy generated by renewable energy sources. Therefore, most of the energy required is produced by fossil fuels. On the other hand, the permanent use of fossil fuels gives rise to global warming, environmental pollution, and depletion of energy resources, which has become one of the most important concerns we have ever encountered. In this context, hydrogen is foreseen as a perfect alternative to fossil fuel for being a constant source of clean and renewable energy [1–5] .

The unique properties of hydrogen, including zero $CO_2$ emissions, high efficiency, and diverse utilization capability, make it a promising energy source [5], [6]. Nevertheless, the future of an economy based on hydrogen is largely limited by finding a cheap, safe, and practical method to store hydrogen at ambient temperature. In general, liquid hydrogen storage and high-pressure gas are two common methods that have not been able to satisfy this demand due to high prices and safety problems [7], [8].

Nanotechnology is a potential gamechanger in this respect. Nanotechnology has proven itself as one of the emerging branches of science with numerous applications in industrial, medicinal, and energy uses [9–12]. Adsorption-based hydrogen storage in nanostructured materials has been attracting a lot of attention in the last years [13–17]. Efforts have been made to assess porous materials such as metal-organic frameworks, zeolites, and covalent organic frameworks for hydrogen storage and purification [18–20]. Those materials show high capacity but usually under extreme operational conditions [21], [22]. Carbon-based materials have been also explored for storing hydrogen, activated carbons, graphene sheets, and carbon nanotubes are some examples [23–27]. Among the different nanomaterials, single-walled carbon nanotubes (SWCNT) have received much attention due to their significant advantages such as large surface area, lightweight, pore structure, and the capability of adsorption at ambient temperature [28–35]. Despite these advantages, the extremely weak Van der Waals interactions between hydrogen and pristine carbon nanotubes limit the adsorption capacity. According to the US Department of Energy (DOE), the target average binding energy between hydrogen and the surface of nanomaterials should be in the range of 0.15-0.6 eV. Materials that have weaker interactions won't capture hydrogen, while stronger interactions would hamper the reversibility of the storage process. Furthermore, an ideal hydrogen storage material should have a gravimetric and volumetric capacity of more than 5.5 wt% and 40 g/L, respectively [36], [37].

To improve hydrogen absorption, surface modification is introduced as one of the most effective methods [38–42]. By modifying the surface of the carbon nanotube (CNT), the binding energy can be dramatically increased leading to high storage capacity. Seenithurai et al. [43] explored Al-decorated carbon nanotube for hydrogen storage applications. According to their findings, the structure adsorbed four hydrogen molecules per aluminum atom, obtaining a storage capacity of 6.15 wt% with an average binding energy of 0.214 eV/$H_2$. Yang et al. [44] performed a surface modification of single-walled carbon nanotubes by dual-Ti doping. They monitored a controlled hydrogen adsorption and observed a dissociation of the first-bounded di-hydrogen as a consequence of the strong chemisorption. Subsequent capture of six hydrogen molecules per Ti atom with an adsorption energy of 0.198 eV/$H_2$ was recorded. Verdinelli et al. [45] reported Ru-doped single-wall carbon nanotubes (SWCNT) with up to five Ru atoms per unit cell without clusterization. The Ru-doped structure showed an adsorption capacity of four hydrogen molecules per Ru atom and adsorption energies of -0.83 eV/$H_2$. Similar findings were reported by Liu et al. [46] obtaining high adsorption capacity, i.e. four di-hydrogen per impurity, in ruthenium and boron-doped carbon nanotubes. The calculated binding energy, ranged from 1.06 and 1.15 eV/$H_2$, which is above the DOE target range, indicating high interaction and hampering the release of the captured hydrogen. Sawant et al. further investigated the effect of boron doping on the hydrogen adsorption properties of carbon nanotubes obtaining gravimetric capacities up to 2.5 wt% for single (5-15 nm diameters) and multi-walled (20-25 nm diameter) CNTs [47]. Reyhani et al. [48] investigated hydrogen storage under ambient conditions in multiwall carbon nanotubes (MWCNT) encapsulating Ca, Co, Fe, Ni, and Pd nanoparticles. A maximum gravimetric capacity of 4.6 wt% was obtained for the Pd-decorated MWCNT of about 60 nm diameter. It is worth noticing that, in most of the mentioned literature, the three representative DOE goals (volumetric capacity, gravimetric capacity, and binding energy) are not reached simultaneously.

The novelty of the present work is the use of gold as an efficient doping element on the surface of carbon nanotubes for hydrogen storage applications. Gold is selected because *i)* gold nanoparticles encapsulation in other nanostructures has demonstrated to enhance the hydrogen adsorption capacity [49–54] *ii)* previous works reported the viability of post-synthetic modification of carbon nanotubes using gold [55–58]. However, the capability for hydrogen storage is still unexplored. To investigate gold-doped carbon nanotubes for hydrogen storage applications, density functional theory has been used as a proven method for exploring the properties of the materials in the ground state.

In this work, the adsorption of hydrogen on Au-doped CNT has been extensively investigated. Moreover, the effect of different CNT diameters on the volumetric and gravimetric capacity of Au-doped CNT has been considered. According to the results, the addition of gold doping increased the average hydrogen binding energy to the DOE target range. This positive effect has been investigated in depth using PDOS, LDOS, and charge transfer. In addition, volumetric capacity, gravimetric capacity, and desorption temperature has been calculated, analyzing the effect of the Au-doped CNT diameter. Simultaneously achieving all DOE goals is rarely possible, as shown in literature [43–48] and Au-doped CNT is not an exception. The results show an underachievement for the gravimetric capacity for the small diameter, while obtaining values close to the desired range as it increases. We obtain outstanding results for volumetric capacity in all doped structures that range from three to six times the DOE target. Based on these findings and the temperature desorption analysis, Au-doped CNT is a suitable candidate for reversible hydrogen storage under ambient operational conditions.

## Calculation methodology

First principle density functional theory (DFT) implemented in the SIESTA package is used to perform the simulations [59]. The exchange-correlation energy is calculated using the generalized gradient approximation (GGA) formulated by Perdew-Burke-Ernzerhof (PBE) [60]. The Troullier-Martins norm-conserving pseudo-potentials are used to describe the potential of atomic cores and related core electrons with valence electrons. The basis wave functions are expanded over a set of numerical atomic orbitals (NAOs) at the level of double-ζ plus polarization (DZP). The Hellmann-Feynman theorem is used for all atomic relaxation coordinates by minimizing the forces on individual atoms down to less than 0.03 eV/Å. Moreover, real space mesh cut-off energy of 150 Ry is chosen. The Brillouin zone is sampled by a 4×4×4 k-points grid under the

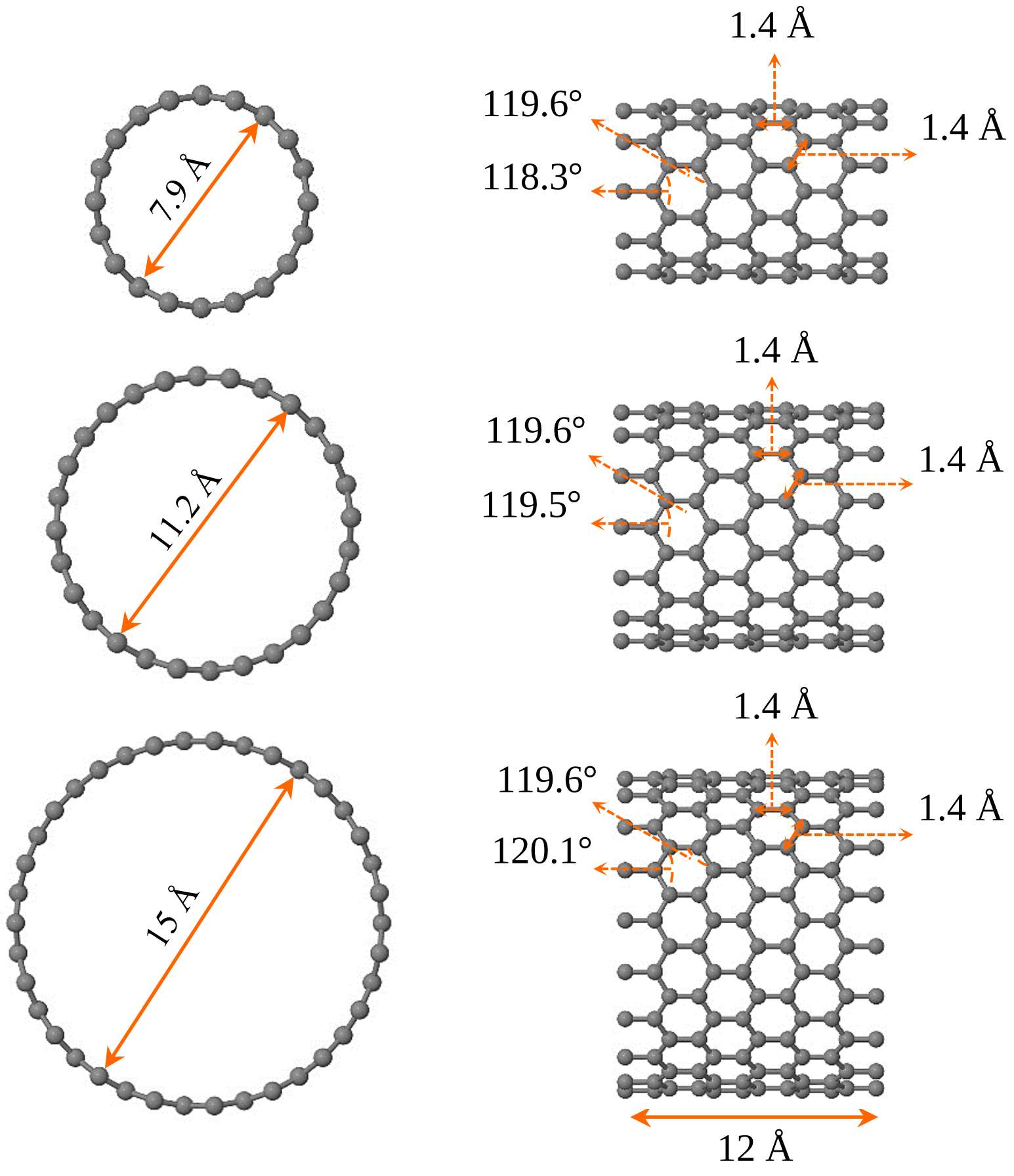

*Figure 1: Schematic representation of carbon nanotube structures. On the right, sideview of the tubes are depicted.*

Monkhorst-pack scheme.

In order to investigate the effect of CNT diameter on hydrogen storage capacity, three zigzag single walled carbon nanotubes with diameters of 7.9, 11.2, and 15 Å and a length of about 12 Å have been considered (Figure 1). The carbon nanotubes contain 120, 168, and 228 carbon atoms per unit cell, and are termed $CNT_{120}$, $CNT_{168}$, $CNT_{228}$, respectively. Crystallographic information is provided as supplementary material. To represent an infinitely long carbon nanotube, the periodic hexagonal supercell has been repeated in lattice parameters of a=25 Å, b=25 Å, and c=12.789 Å.

All nomenclature and parameters are collected and defined in the glossary, table S1 in the Supporting Information (SI).

The binding ($E_b$) and formation energies ($E_f$) of the gold atom (Au) have been determined using Eqs. (1) and (2):

$$E_b\,(Au) = \left[\frac{E_{tot(CNT)} + E_{tot(nAu)} - E_{tot(nAu-CNT)}}{n}\right], (1)$$

$$E_f = E_{tot(1Au-CNT)} - \frac{E_{tot(C(graph))}}{m} - \frac{E_{tot(Au(bulk))}}{n}, (2)$$

where $E_{tot(CNT)}$ represents the total energy of the carbon nanotube per unit cell, $E_{tot(nAu)}$ denotes the total energy of *n* gold atoms, $E_{tot(nAu-CNT)}$ indicates the total energy of the carbon nanotube doped with *n* gold atoms per unit cell, *n* being the number of gold atoms on the surface of the carbon nanotube. In Eq. (2), $E_{tot(1Au-CNT)}$ is the total energy of the Au-doped CNT, $E_{tot(C(graph))}$ is the total energy of the most stable structure of carbon (graphene) with m carbon atoms, and $E_{tot(Au(bulk))}$ is the total energy of the most stable nanostructure of gold (bulk) with n gold atoms.

The average binding energy per hydrogen molecule has been calculated using Eq. (3):

$$E_b\,(H_2) = \left[\frac{E_{tot(nAu-CNT)} + E_{tot(iH_2)} - E_{tot((nAu-CNT)+iH_2)}}{i}\right], (3)$$

where $E_{tot(nAu-CNT)}$ is the total energy of the host structure per unit cell, $E_{tot(iH_2)}$ denotes the total energy of *i* hydrogen molecules, $E_{tot((nAu-CNT)+iH_2)}$ indicates the total energy of the *i* adsorbed hydrogen molecules on the surface of the carbon nanotube doped with *n* gold atoms per unit cell.

Mulliken charge transfer analysis is used to calculate the charge transfer between hydrogen molecules and the surface of *nAu-CNT*:

$$q_t = q_{(ads-H_2)} - q_{(iso-H_2)}, (4)$$

where $q_{(ads-H_2)}$ indicates the total charge of the adsorbed hydrogen molecule and $q_{(iso-H_2)}$ is the total charge of the isolated hydrogen molecule.

Moreover, Van 't Hoff's equation is used to calculate the desorption temperature of hydrogen molecules on the surface of *nAu-CNT* [5]:

$$T_d = \frac{E_{ads}}{k_B}\left(\frac{\Delta S}{R} - \ln p\right)^{-1}, (5)$$

where $E_{ads}$, $k_B$, $\Delta S$, R, and $p$ represent the average adsorption energy per hydrogen molecule, the Boltzmann constant, the change in hydrogen entropy from the gas to the liquid phase, the gas constant, and the atmospheric pressure, respectively .

In addition, the following equation is applied to compute gravimetric storage capacity of *nAu-CNT*:

$$wt\,\% = \frac{i\,W_{H_2}}{i\,W_{H_2} + W_{(nAu-CNT)}} \times 100, (6)$$

here $i$ is the number of adsorbed hydrogen molecules, $W_{H_2}$the molecular weight of the hydrogen molecule, and $W_{(nAu-CNT)}$ is the molecular weight of the *nAu-CNT*.

Finally, the volumetric storage capacity was calculated using:

$$n_v = \left(\frac{i\;W_{H_2}\;1000\;\rho_{crystal}}{W_{(nAu-CNT)}}\right), (7)$$

$$\rho_{crystal} = \left(\frac{Z\;M_V}{V_C\;N_A}\right), (8)$$

Z, $M_V$, $V_C$, and $N_A$ represent the number of atoms per unit cell, the molecular weight of atoms per unit cell (g/mol), the volume of the unit cell ($cm^3$), and the Avogadro constant, respectively.

## Results and discussions

In this work, carbon nanotubes were studied as a promising substrate for hydrogen storage applications due to their unique properties. Initially, a hydrogen molecule is placed near the pristine carbon nanotube surface in different configurations and orientations to determine the affinity. The results show binding energies ranging from 0.07 to 0.15 eV in the external surface and inside the pore for $CNT_{120}$. All calculated binding energies for different hydrogen configurations inside and outside *nAu-CNTs* can be found in Table S2 in the SI. Although, the binding energy inside the carbon nanotube is higher than the external surface, it is in the lower limit of the DOE range (0.15-0.6 eV). The weak interactions indicate that pristine carbon nanotube material is not a suitable candidate for hydrogen storage and the determination of parameters such as volumetric and gravimetric capacity become irrelevant. To increase the adsorption capacity of carbon nanotubes, surface modification is used. The addition of gold impurities serves the purpose of creating new adsorption centers, increasing the external surface-hydrogen interaction. We take the limits of the DOE target binding energy to ensure reversibility of the adsorption-desorption process.

In order to identify the most stable structure of a Au-doped carbon nanotube, gold has been placed in three different configurations: *i)* on top of the carbon atom, *ii)* between the carbon bonds, and *iii)* in the center of the hexagon. The schematic representation of the Au training positions can be found in Figure S1 in the supporting information (SI). The formation and binding energies for the configurations under study have been calculated using Eqs. (1) and (2). Two configurations of Au-doped carbon nanotubes are identified as the most stable ones obtaining similar energies with minima in configuration *ii* ($E_f$ = -0.936, $E_b$ = 0.73) and *iii* ($E_f$ = -0.928, $E_b$ = 0.72).

Figure 2 indicates the partial density of states (PDOS) associated with the change in the gold orbitals. Orbitals *s* and *d* of gold atom are analyzed after being adsorbed in the carbon nanotube surface and compared with the isolated gold atom. As shown in Figure 2, after adding gold to the surface of carbon nanotubes, the *s* and *d*-orbitals of gold have changed dramatically. These orbitals have shifted to the left (lower energies). The gold atom acquired charge from the surface of the carbon nanotube, increasing its stability. It can also be observed that the number of available states per energy decreased, i.e., some empty gold orbitals have been occupied due to the acquiring of charge from the surface of the carbon nanotube. These diagrams are extremely useful to understand the effect of gold doping on the hydrogen-CNT binding energy.

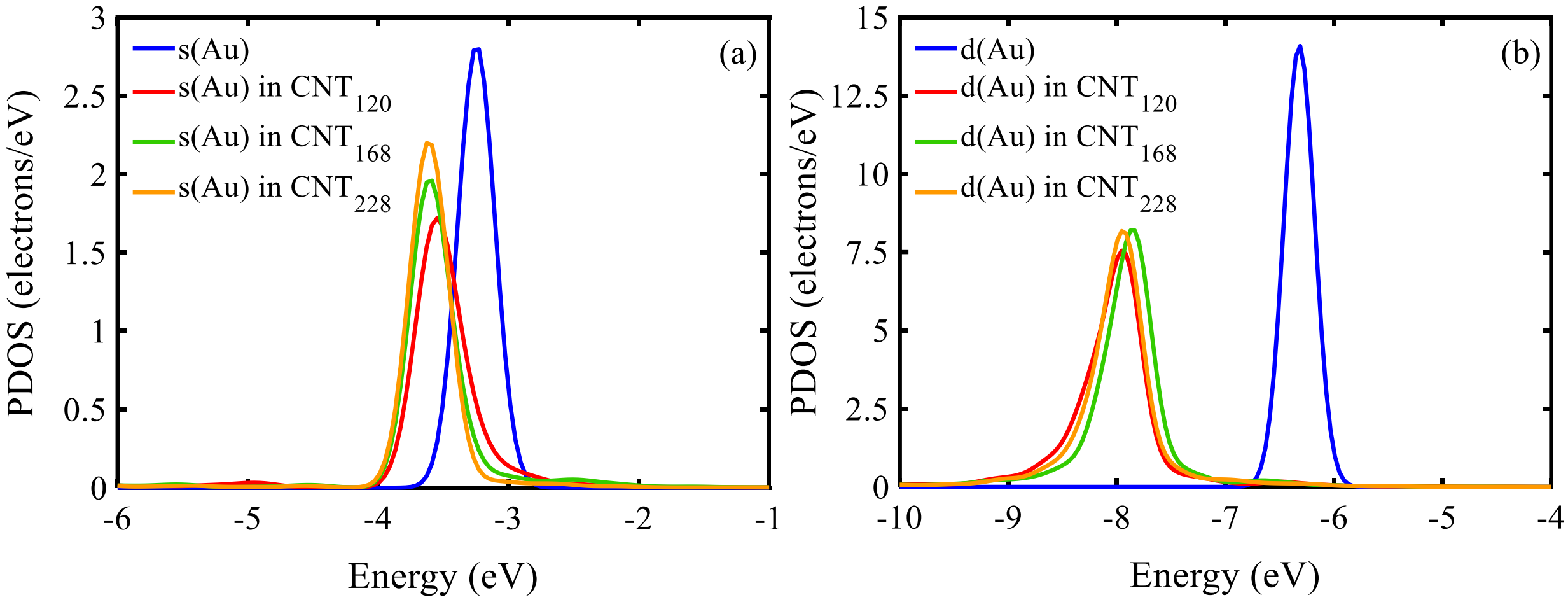

*Figure 2: PDOS of (a) s and (b) d-orbitals of the pure Au atom and Au atom on the surface of the CNT with different diameters.*

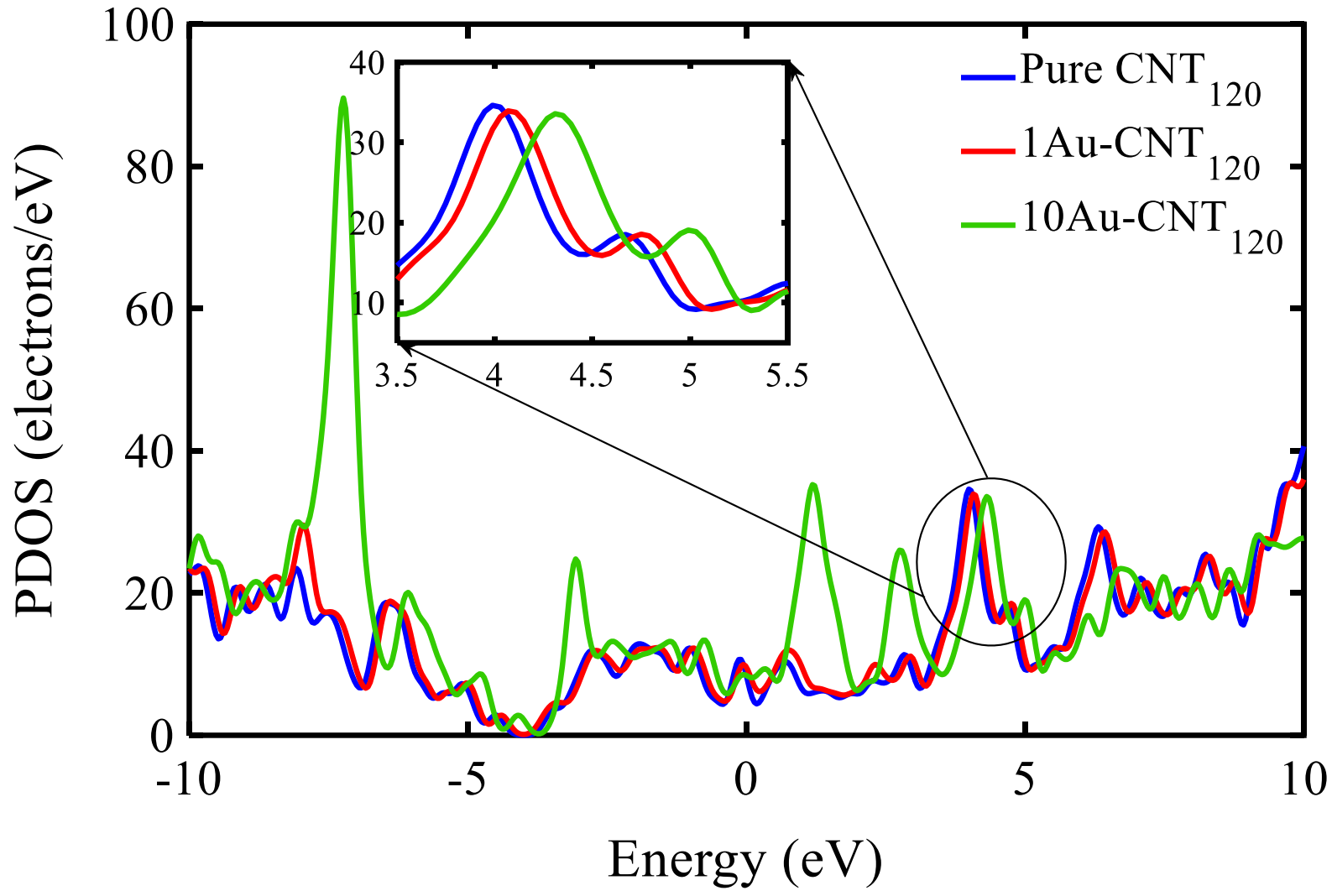

*Figure 4: Total PDOS of pristine and nAu-CNT$_{120}$. The surface is saturated for n=10.*

To analyze the effect of gold impurities on the structural properties of carbon nanotubes, the PDOS were calculated at two conditions, doped (*1Au-CNTs*) and Au-saturated CNTs. Au-saturated CNTs have the maximum number of gold impurities *n* per unit cell so that the entire external surface of the CNT is covered without cluster formation (*nAu-CNT*). Figure 3 shows the PDOS of the pristine, doped, and saturated CNT$_{120}$. As shown in the figure, the addition of a single impurity produces a slight shift in energies to the right. The displacement of the energy becomes more important on the saturated structure. The shift in energy is attributed to the charge transfer between the surface of the carbon nanotube and gold atoms. As a result, the addition of gold atoms to the surface of the carbon nanotube causes a decrease in structural stability significantly increasing the number of available states. The shown instability and increase of available states is translated into a predisposition of the doped structures to engage in physio-chemical reactions. The same behavior is also found for *n*Au-CNT$_{168}$ and *nAu-CNT$_{228}$*. All PDOS diagrams are shown in Figure S2 in the SI.

Charge transfer has been calculated by the Mulliken charge analysis to support the results and conclusions from the gold and structure PDOSs. In general, the negative amount of charge transfer is defined as the charge transferred from the surface of the material to the adsorbate. Here it is also used for the charge transferred between the structure and the gold atoms. For *nAu-CNTs*, the charge transfers from the surface of the carbon nanotube to the gold impurity causing an increase of the stability of the gold impurity while the carbon nanotube shows displacement towards instability. This conclusion is supported by the results of the PDOSs (Figures 2 and 3). The charge transferred is expected to increase with the number of gold impurities added to the surface. The CNTs doped with a single gold particle show a charge transfer of -0.184e, -0.153e, and -0.184e for $CNT_{120}$, $CNT_{168}$, and $CNT_{228}$, respectively. The Au-saturated structures show -1.07e, -1.48e, and -1.72e. As all Au-saturated CNTs have C/Au ratios of about 12, it is clear that the increase of diameter has an effect on the charge transfer. These results are in good agreement with the creation of new available states and the shift of the PDOS diagrams to the left after the addition of gold atoms.

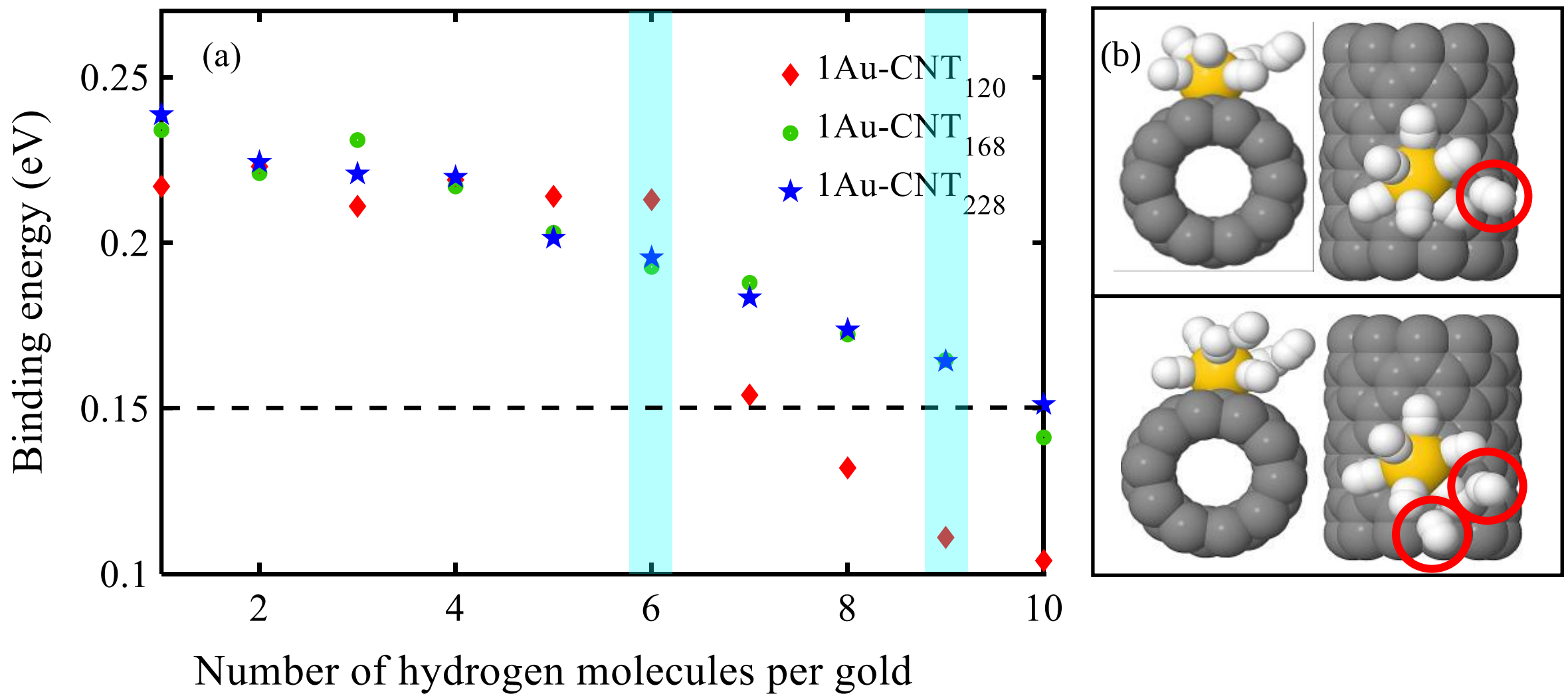

*Figure 3: Binding energy of hydrogen adsorption on 1Au-CNTs with different diameters (dashed line: the lower boundary of binding energy as defined by DOE).*

The binding energy between hydrogen molecules and *nAu-CNT* is calculated to explore the effect of gold on the hydrogen storage capacity of carbon nanotubes. Figure 4a shows the binding energy per hydrogen molecule in *1Au-CNTs* when increasing the number of adsorbed hydrogen molecules from one to ten. In the small structure, $CNT_{120}$, the binding energy fluctuates around 0.21 eV from one to six hydrogen molecules. From that point on, the adsorption of new hydrogen molecules causes a dramatic decrease in the binding energy to values below the DOE target. This behaviour is caused by the non-bonded (weakely adsorbed) hydrogen molecules as highlighted in Figure 4b. $1Au\text{-}CNT_{168}$ and $1Au\text{-}CNT_{228}$ show a different behaviour. The calculated binding energy of a single molecule is higher than for the small structure but monotonously decreases with the adsorption of more molecules. These structures can absorb a maximum of nine hydrogen molecules per gold atom, maintaining binding energies above the limit of the DOE taget. Therefore, the increase of diameter shows a possitive effect on the hydrogen storage capacity of *nAu-CNTs*. All calculated binding energies represented in Figure 4 can be found in Table S3 in the SI.

Local Density of State (LDOS) is used to analyze the charge distribution after the addition of gold impurities. LDOS is a function of energy and position that shows the weighted DOS by the amplitude of the implied wave functions at different points in space [61] As shown in Figure 5, gold impurities create new states around them (5b and 5c). The effect of the newly available states, used to adsorb hydrogen, is depicted in Figure 5d. The effect of the adsorption of hydrogen in the LDOS representations is in agreement with the PDOS results (Figure 6).

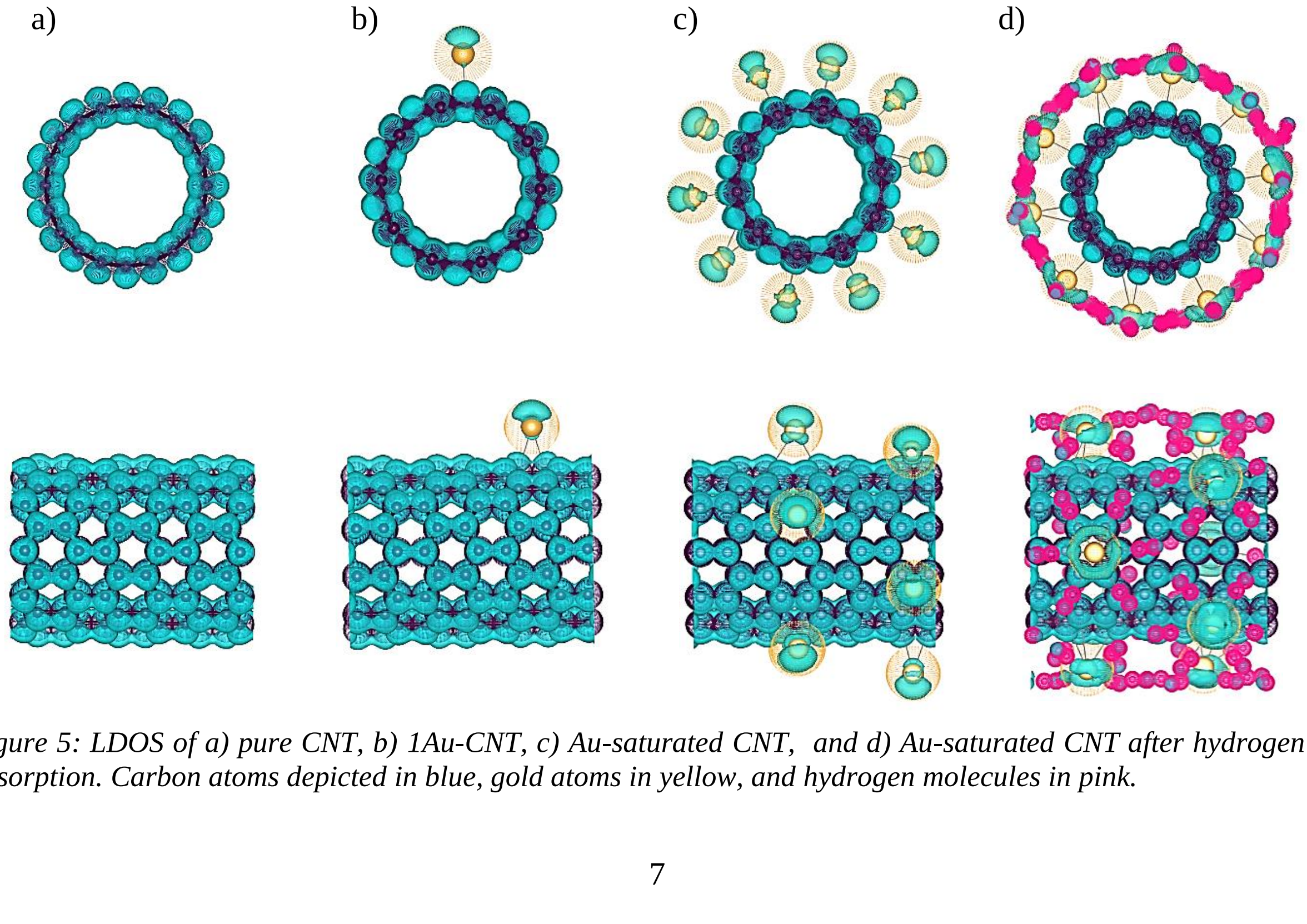

*Figure 5: LDOS of a) pure CNT, b) 1Au-CNT, c) Au-saturated CNT, and d) Au-saturated CNT after hydrogen adsorption. Carbon atoms depicted in blue, gold atoms in yellow, and hydrogen molecules in pink.*

Figure 6 shows the PDOS of the Au-saturated $CNT_{120}$ before and after the adsorption of hydrogen using the pristine structure as reference. As shown before (Figure 3) the addition of gold impurities causes an energy shift to the right, creating structure instability. The new available states and higher energy values of the Au-saturated CNTs enhance the capability of the structures to absorb external species. After the adsorption of six molecules of hydrogen per gold atom in *10Au-CNT*$_{120}$, the energy is shifted again to the right, increasing the system stability, and the available states are reduced to similar values as for the pristine CNT. The Fermi energy also shows this behavior, i.e. increase of energy caused by the doping, and regaining stability after the adsorption of hydrogen. The total PDOS diagram of *nAu-CNT*$_{168}$ and *nAu-CNT*$_{228}$ show similar behavior  and can be found in Figure S3 in the SI.

The charge transfer was calculated using the Mulliken method. For *10Au-CNT*$_{120}$, a charge transfer of 1.37e was calculated after the adsorption of six molecules of hydrogen per gold atom. Therefore the charge is transferred from the adsorbed hydrogen to the surface of the Au-saturated CNT. The obtained values for *14Au-CNT*$_{168}$ and *18Au-CNT*$_{228}$ increased to 2.69e and 3.99e, respectively. These results are consistent with PDOS results (Figure 6 and S3).

Since the binding energy of hydrogen with the *nAu-CNTs* is found to be in the DOE target range, the next step is to determine the gravimetric and volumetric capacities for the carbon nanotubes. In the case of $CNT_{120}$, six hydrogen molecules per gold atom were adsorbed. After saturating the structure in a configuration that prevents gold cluster formation, gravimetric and volumetric capacities of 3.42 wt% and 155.44 g/l are found. The obtained gravimetric capacity is below 5.5 wt% (DOE target for 2025), while the results show high volumetric capacity (>> DOE target of 40 g/L). To analyze the effect of the diameter for hydrogen storage $CNT_{168}$ and $CNT_{228}$ were doped to maximum capacity (C/Au ratio of 12). The results show an increase of gravimetric and volumetric capacity for both, reaching values of 5.05 wt% and 297.62 g/l for *14Au-CNT*$_{168}$ and 4.94 wt% and 333.08 g/l for *18Au-CNT*$_{228}$. The increase of  the diameter of carbon nanotubes enhances the volumetric capacity surpassing the DOE target, predicting high performance of the structures under study for hydrogen storage applications.

To assess the performance of the doped CNTs for practical applications, the desorption temperature is calculated using Eq. (5) and the obtained data can be found in Table S4 of the SI. Figure 7 shows the desorption temperature as a function of the number of hydrogen molecules adsorbed per gold impurity. The desorption temperature of *1Au-CNT$_{120}$* shows a plateau from one to six adsorbed hydrogen molecules per gold impurity that fluctuates around 275 K. Corresponding to the dissociation showed in Figure 4b, the increase of hydrogen from that point on causes a decrease in the desorption temperature, from 200 to 125 K. *1Au-CNT$_{168}$* and *1Au-CNT$_{228}$* show different behavior. These doped carbon nanotubes have a continuous decrease of desorption temperature with increasing numbers of adsorbed hydrogen molecules going from 300 to 210 K at maximum capacity per gold impurity (highlighted in cyan).

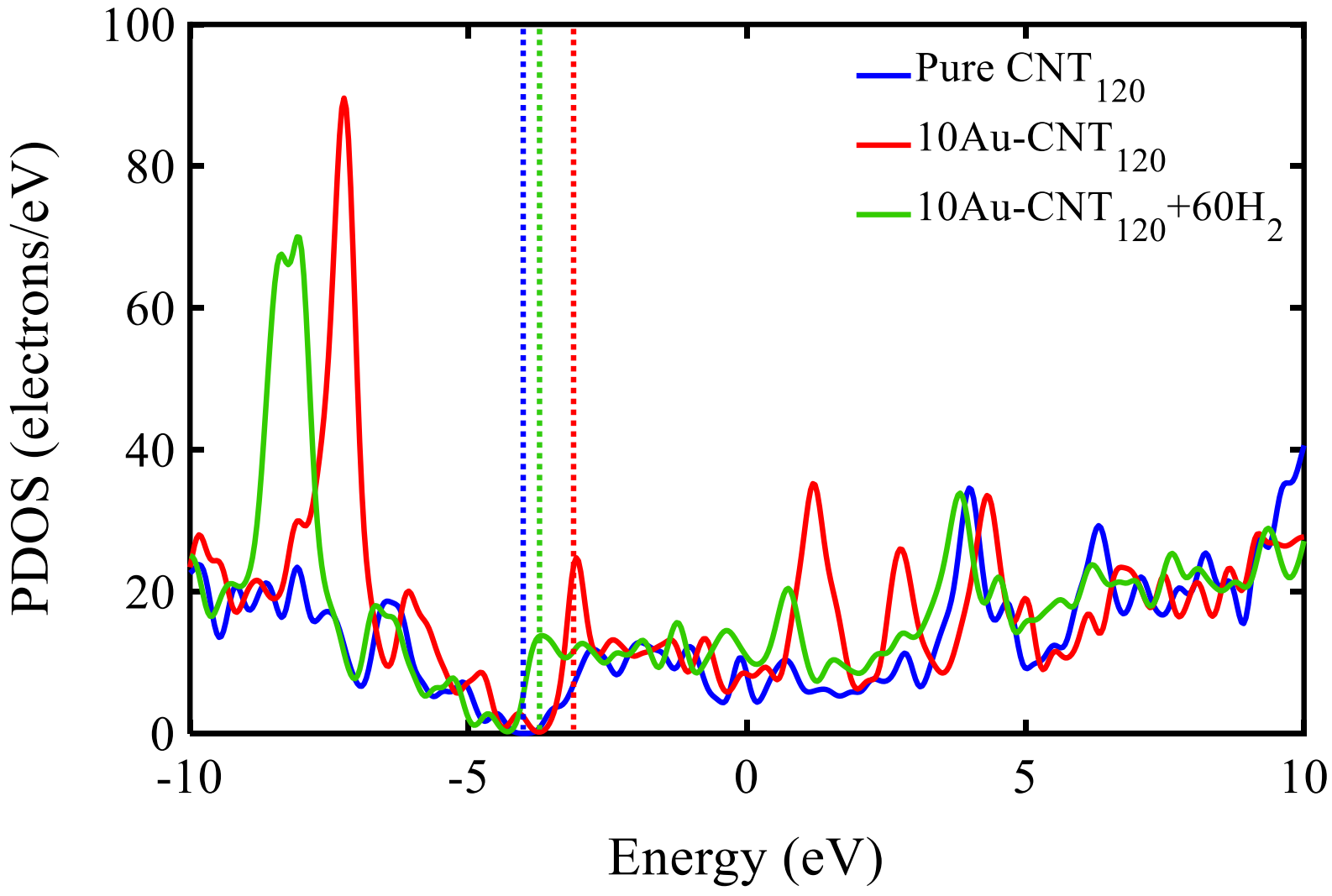

*Figure 6: PDOS of carbon nanotube, pristine (blue), saturated with gold (red), after hydrogen adsorption in the saturated structure (green). Dashed lines indicate the Fermi Energy ($E_F$) per configuration.*

To further assess the doped *nAu-CNT* structures, the desorption temperature of the gold-saturated structures at maximum hydrogen capacity per gold atom was also calculated (inset Figure 7). The results show desorption temperatures of 278, 207, and 209 K (*10Au-CNT$_{120}$*, *14Au-CNT$_{168}$*, and *18Au-CNT$_{228}$*) to achieve complete regeneration of the structures, i.e., full hydrogen release.

These results prove that Au-doped CNTs can act as a reversible hydrogen storage substrate with at relatively high desorption temperature (above 200 K).

## Conclusions

In this paper, the effect of doping carbon nanotubes (in three different diameters) with gold on hydrogen storage properties has been investigated using density functional theory (DFT). The calculation results show that the addition of gold impurities to the surface of carbon nanotubes creates empty states on the surface of this structure that can be filled by electrons and enhances the adsorption of hydrogen. The Au-doped CNTs became instable, prompting these structures to adsorb hydrogen and reach stability. Binding energy, partial density of states (PDOS), and charge transfer computations have been performed to gain an in-depth understanding of the hydrogen adsorption mechanism in the Au-doped CNTs. The results indicate that six and nine hydrogens per gold atom can be absorbed by CNT$_{120}$ and CNT$_{168, 228}$. Gravimetric and volumetric capacities have been calculated to be about 3.42 wt% and 154.44 g/l for CNT$_{120}$, 5.049 wt% and 297.62 g/l for CNT$_{168}$, and 4.93 wt% and 333.08 g/l for CNT$_{228}$. The volumetric capacity results are much higher than the DOE target for 2025, especially in the case of CNT$_{168}$ and CNT$_{228}$. The results confirmed the positive effect of increasing CNT diameter on increasing the number of adsorbed hydrogen molecules. Moreover, the desorption temperature for these structures is predicted to be in the range of 200-

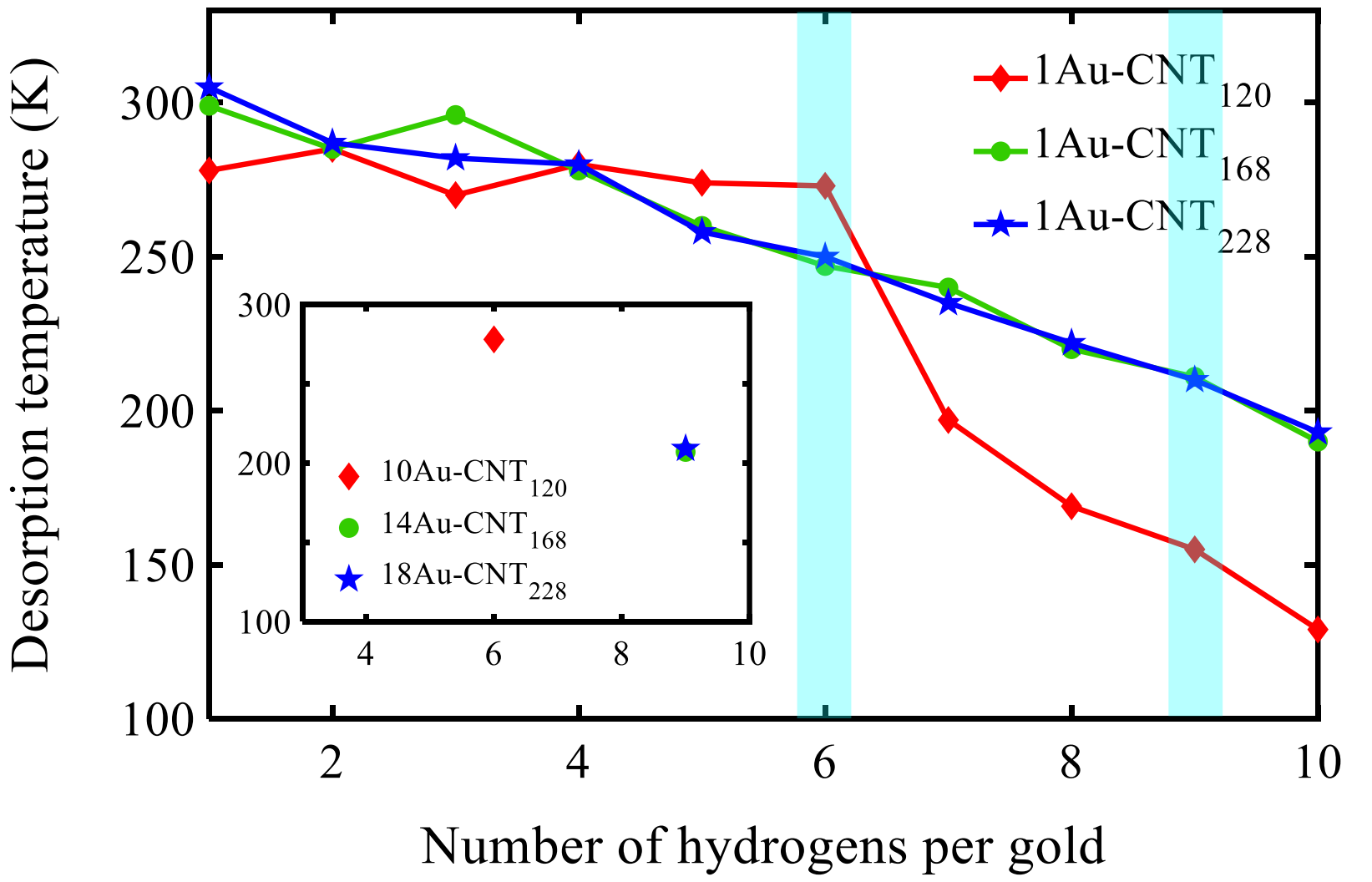

*Figure 7: Desorption temperature of 1Au-CNTs with different diameters. The inset shows the desorption temperatures of saturated CNTs.*

300 K. We conclude that Au-doped CNT can be an appropriate candidate for hydrogen storage applications.

## Conflicts of interest

The authors declare no conflicts of interest.

## Acknowledgment

EIRES – Eindhoven Institute for Renewable Energy Systems.

## References

[1] D. Parra, L. Valverde, F. J. Pino, and M. K. Patel, "A review on the role, cost and value of hydrogen energy systems for deep decarbonisation," *Renewable and Sustainable Energy Reviews*, vol. 101. Elsevier Ltd, pp. 279–294, Mar. 01, 2019. doi: 10.1016/j.rser.2018.11.010.

[2] J. O. Abe, A. P. I. Popoola, E. Ajenifuja, and O. M. Popoola, "Hydrogen energy, economy and storage: Review and recommendation," *International Journal of Hydrogen Energy*, vol. 44, no. 29. Elsevier Ltd, pp. 15072–15086, Jun. 07, 2019. doi: 10.1016/j.ijhydene.2019.04.068.

[3] S. Sharma, S. Basu, N. P. Shetti, and T. M. Aminabhavi, "Waste-to-energy nexus for circular economy and environmental protection: Recent trends in hydrogen energy," *Science of the Total Environment*, vol. 713, Apr. 2020, doi: 10.1016/j.scitotenv.2020.136633.

[4] Y. Zhou, R. Li, Z. Lv, J. Liu, H. Zhou, and C. Xu, "Green hydrogen: A promising way to the carbon-free society," *Chinese Journal of Chemical Engineering*, vol. 43. Materials China, pp. 2–13, Mar. 01, 2022. doi: 10.1016/j.cjche.2022.02.001.

[5] S. Hu, Y. Yong, Z. Zhao, R. Gao, Q. Zhou, and Y. Kuang, "C7N6 monolayer as high capacity and reversible hydrogen storage media: A DFT study," *Int J Hydrogen Energy*, vol. 46, no. 42, pp. 21994–22003, Jun. 2021, doi: 10.1016/j.ijhydene.2021.04.053.

[6] F. Emir and F. V. Bekun, "Energy intensity, carbon emissions, renewable energy, and economic growth nexus: New insights from Romania," *Energy and Environment*, vol. 30, no. 3, pp. 427–443, May 2019, doi: 10.1177/0958305X18793108.

[7] A. Züttel, "Hydrogen storage methods," *Naturwissenschaften*, vol. 91, no. 4. pp. 157–172, Apr. 2004. doi: 10.1007/s00114-004-0516-x.

[8] S. Niaz, T. Manzoor, and A. H. Pandith, "Hydrogen storage: Materials, methods and perspectives," *Renewable and Sustainable Energy Reviews*, vol. 50. Elsevier Ltd, pp. 457–469, May 30, 2015. doi: 10.1016/j.rser.2015.05.011.

[9] M. H. Ahmadi *et al.*, "Renewable energy harvesting with the application of nanotechnology: A review," *International Journal of Energy Research*, vol. 43, no. 4. John Wiley and Sons Ltd, pp. 1387–1410, Mar. 25, 2019. doi: 10.1002/er.4282.

[10] M. A. Shah, B. M. Pirzada, G. Price, A. L. Shibiru, and A. Qurashi, "Applications of nanotechnology in smart textile industry: A critical review," *Journal of Advanced Research*, vol. 38. Elsevier B.V., pp. 55–75, May 01, 2022. doi: 10.1016/j.jare.2022.01.008.

[11] M. Zain *et al.*, "Applications of nanotechnology in biological systems and medicine," in *Nanotechnology for Hematology, Blood Transfusion, and Artificial Blood*, Elsevier, 2021, pp. 215–235. doi: 10.1016/B978-0-12-823971-1.00019-2.

[12] N. Dey *et al.*, "Nanotechnology-assisted production of value-added biopotent energy-yielding products from lignocellulosic biomass refinery – A review," *Bioresource Technology*, vol. 344. Elsevier Ltd, Jan. 01, 2022. doi: 10.1016/j.biortech.2021.126171.

[13] A. Ahmed *et al.*, "Exceptional hydrogen storage achieved by screening nearly half a million metal-organic frameworks," *Nat Commun*, vol. 10, no. 1, Dec. 2019, doi: 10.1038/s41467-019-09365-w.

[14] P. Ramirez-Vidal *et al.*, "A Step Forward in Understanding the Hydrogen Adsorption and Compression on Activated Carbons," *ACS Appl Mater Interfaces*, vol. 13, no. 10, pp.

12562–12574, Mar. 2021, doi: 10.1021/acsami.0c22192.

[15] S. Cao, B. Li, R. Zhu, and H. Pang, "Design and synthesis of covalent organic frameworks towards energy and environment fields," *Chemical Engineering Journal*, vol. 355. Elsevier B.V., pp. 602–623, Jan. 01, 2019. doi: 10.1016/j.cej.2018.08.184.

[16] M. Hirscher *et al.*, "Materials for hydrogen-based energy storage – past, recent progress and future outlook," *J Alloys Compd*, vol. 827, Jun. 2020, doi: 10.1016/j.jallcom.2019.153548.

[17] R. Kumar, A. Khan, A. M. Asiri, and H. Dzudzevic-Cancar, "Preparation methods of hydrogen storage materials and nanomaterials," in *Nanomaterials for Hydrogen Storage Applications*, Elsevier, 2021, pp. 1–16. doi: 10.1016/b978-0-12-819476-8.00002-5.

[18] S. H. Sang, H. Furukawa, O. M. Yaghi, and W. A. Goddard, "Covalent organic frameworks as exceptional hydrogen storage materials," *J Am Chem Soc*, vol. 130, no. 35, pp. 11580–11581, Sep. 2008, doi: 10.1021/ja803247y.

[19] K. Suresh, D. Aulakh, J. Purewal, D. J. Siegel, M. Veenstra, and A. J. Matzger, "Optimizing Hydrogen Storage in MOFs through Engineering of Crystal Morphology and Control of Crystal Size," *J Am Chem Soc*, vol. 143, no. 28, pp. 10727–10734, Jul. 2021, doi: 10.1021/jacs.1c04926.

[20] M. Erdős *et al.*, "In silico screening of zeolites for high-pressure hydrogen drying," *ACS Appl Mater Interfaces*, vol. 13, no. 7, pp. 8383–8394, Feb. 2021, doi: 10.1021/acsami.0c20892.

[21] J. Ren, N. M. Musyoka, H. W. Langmi, M. Mathe, and S. Liao, "Current research trends and perspectives on materials-based hydrogen storage solutions: A critical review," *International Journal of Hydrogen Energy*, vol. 42, no. 1. Elsevier Ltd, pp. 289–311, Jan. 05, 2017. doi: 10.1016/j.ijhydene.2016.11.195.

[22] N. P. Stadie, J. J. Vajo, R. W. Cumberland, A. A. Wilson, C. C. Ahn, and B. Fultz, "Zeolite-templated carbon materials for high-pressure hydrogen storage," *Langmuir*, vol. 28, no. 26, pp. 10057–10063, Jul. 2012, doi: 10.1021/la302050m.

[23] Q. Hu, Y. Lu, and G. P. Meisner, "Preparation of nanoporous carbon particles and their cryogenic hydrogen storage capacities," *Journal of Physical Chemistry C*, vol. 112, no. 5, pp. 1516–1523, Feb. 2008, doi: 10.1021/jp076409t.

[24] C. da Wu, T. H. Fang, and J. Y. Lo, "Effects of pressure, temperature, and geometric structure of pillared graphene on hydrogen storage capacity," *Int J Hydrogen Energy*, vol. 37, no. 19, pp. 14211–14216, Oct. 2012, doi: 10.1016/j.ijhydene.2012.07.040.

[25] Z. Jin *et al.*, "Nano-engineered spacing in graphene sheets for hydrogen storage," *Chemistry of Materials*, vol. 23, no. 4, pp. 923–925, Feb. 2011, doi: 10.1021/cm1025188.

[26] X. Peng, J. M. Vicent-Luna, and Q. Jin, "Water-Gas Shift Reaction to Capture Carbon Dioxide and Separate Hydrogen on Single-Walled Carbon Nanotubes," *ACS Appl Mater Interfaces*, vol. 13, no. 9, pp. 11026–11038, Mar. 2021, doi: 10.1021/acsami.1c00145.

[27] V. Jiménez, A. Ramírez-Lucas, P. Sánchez, J. L. Valverde, and A. Romero, "Hydrogen storage in different carbon materials: Influence of the porosity development by chemical activation," *Appl Surf Sci*, vol. 258, no. 7, pp. 2498–2509, Jan. 2012, doi: 10.1016/j.apsusc.2011.10.080.

[28] M. Mohan, V. K. Sharma, E. A. Kumar, and V. Gayathri, "Hydrogen storage in carbon materials—A review," *Energy Storage*, vol. 1, no. 2, p. e35, Apr. 2019, doi: 10.1002/est2.35.

[29] E. Boateng and A. Chen, "Recent advances in nanomaterial-based solid-state hydrogen storage," *Mater Today Adv*, vol. 6, Jun. 2020, doi: 10.1016/j.mtadv.2019.100022.

[30] K. Kajiwara, H. Sugime, S. Noda, and N. Hanada, "Fast and stable hydrogen storage in the porous composite of MgH2 with Nb2O5 catalyst and carbon nanotube," *J Alloys Compd*, vol. 893, Feb. 2022, doi: 10.1016/j.jallcom.2021.162206.

[31] S. ullah Rather, "Preparation, characterization and hydrogen storage studies

of carbon nanotubes and their composites: A review," *International Journal of Hydrogen Energy*, vol. 45, no. 7. Elsevier Ltd, pp. 4653–4672, Feb. 07, 2020. doi: 10.1016/j.ijhydene.2019.12.055.

[32] M. Liu *et al.*, "Novel 1D carbon nanotubes uniformly wrapped nanoscale MgH2 for efficient hydrogen storage cycling performances with extreme high gravimetric and volumetric capacities," *Nano Energy*, vol. 61, pp. 540–549, Jul. 2019, doi: 10.1016/j.nanoen.2019.04.094.

[33] J. Yuan *et al.*, "Ni-Doped Carbon Nanotube-Mg(BH4)2Composites for Hydrogen Storage," *ACS Appl Nano Mater*, vol. 4, no. 2, pp. 1604–1612, Feb. 2021, doi: 10.1021/acsanm.0c02738.

[34] J. F. N. Dethan and V. Swamy, "Mechanical and thermal properties of carbon nanotubes and boron nitride nanotubes for fuel cells and hydrogen storage applications: A comparative review of molecular dynamics studies," *International Journal of Hydrogen Energy*, vol. 47, no. 59. Elsevier Ltd, pp. 24916–24944, Jul. 12, 2022. doi: 10.1016/j.ijhydene.2022.05.240.

[35] S. I. Hwang *et al.*, "Metal-Organic Frameworks on Palladium Nanoparticle-Functionalized Carbon Nanotubes for Monitoring Hydrogen Storage," *ACS Appl Nano Mater*, Oct. 2022, doi: 10.1021/acsanm.2c00998.

[36] H. W. Langmi, N. Engelbrecht, P. M. Modisha, and D. Bessarabov, "Hydrogen storage," in *Electrochemical Power Sources: Fundamentals, Systems, and Applications*, Elsevier, 2022, pp. 455–486. doi: 10.1016/B978-0-12-819424-9.00006-9.

[37] S. J. Mahdizadeh and E. K. Goharshadi, "Hydrogen storage on graphitic carbon nitride and its palladium nanocomposites: A multiscale computational approach," *Int J Hydrogen Energy*, vol. 44, no. 16, pp. 8325–8340, Mar. 2019, doi: 10.1016/j.ijhydene.2019.02.071.

[38] S. v. Sawant, A. W. Patwardhan, J. B. Joshi, and K. Dasgupta, "Boron doped carbon nanotubes: Synthesis, characterization and emerging applications – A review," *Chemical Engineering Journal*, vol. 427. Elsevier B.V., Jan. 01, 2022. doi: 10.1016/j.cej.2021.131616.

[39] R. S. Rajaura *et al.*, "Structural and surface modification of carbon nanotubes for enhanced hydrogen storage density," *Nano-Structures and Nano-Objects*, vol. 14, pp. 57–65, Apr. 2018, doi: 10.1016/j.nanoso.2018.01.005.

[40] A. Muhulet, F. Miculescu, S. I. Voicu, F. Schütt, V. K. Thakur, and Y. K. Mishra, "Fundamentals and scopes of doped carbon nanotubes towards energy and biosensing applications," *Materials Today Energy*, vol. 9. Elsevier Ltd, pp. 154–186, Sep. 01, 2018. doi: 10.1016/j.mtener.2018.05.002.

[41] D. J. Durbin and C. Malardier-Jugroot, "Review of hydrogen storage techniques for on board vehicle applications," *International Journal of Hydrogen Energy*, vol. 38, no. 34. pp. 14595–14617, Nov. 13, 2013. doi: 10.1016/j.ijhydene.2013.07.058.

[42] J. P. Paraknowitsch and A. Thomas, "Doping carbons beyond nitrogen: An overview of advanced heteroatom doped carbons with boron, sulphur and phosphorus for energy applications," *Energy and Environmental Science*, vol. 6, no. 10. pp. 2839–2855, Oct. 2013. doi: 10.1039/c3ee41444b.

[43] S. Seenithurai, R. K. Pandyan, S. V. Kumar, C. Saranya, and M. Mahendran, "Al-decorated carbon nanotube as the molecular hydrogen storage medium," *Int J Hydrogen Energy*, vol. 39, no. 23, pp. 11990–11998, Aug. 2014, doi: 10.1016/j.ijhydene.2014.05.184.

[44] L. Yang, L. L. Yu, H. W. Wei, W. Q. Li, X. Zhou, and W. Q. Tian, "Hydrogen storage of dual-Ti-doped single-walled carbon nanotubes," *Int J Hydrogen Energy*, vol. 44, no. 5, pp. 2960–2975, Jan. 2019, doi: 10.1016/j.ijhydene.2018.12.028.

[45] V. Verdinelli, A. Juan, and E. German, "Ruthenium decorated single walled carbon nanotube for molecular hydrogen storage: A first-principle study," *Int J Hydrogen Energy*, vol. 44, no. 16, pp. 8376–8383, Mar. 2019, doi: 10.1016/j.ijhydene.2019.02.004.

[46] P. Liu, J. Liang, R. Xue, Q. Du, and M. Jiang, "Ruthenium decorated boron-doped carbon nanotube for hydrogen storage: A first-principle study," *Int J Hydrogen Energy*, vol.

44, no. 51, pp. 27853–27861, Oct. 2019, doi: 10.1016/j.ijhydene.2019.09.019.

[47] S. v. Sawant, S. Banerjee, A. W. Patwardhan, J. B. Joshi, and K. Dasgupta, "Effect of in-situ boron doping on hydrogen adsorption properties of carbon nanotubes," *Int J Hydrogen Energy*, vol. 44, no. 33, pp. 18193–18204, Jul. 2019, doi: 10.1016/j.ijhydene.2019.05.029.

[48] A. Reyhani, S. Z. Mortazavi, S. Mirershadi, A. Z. Moshfegh, P. Parvin, and A. N. Golikand, "Hydrogen storage in decorated multiwalled carbon nanotubes by Ca, Co, Fe, Ni, and Pd nanoparticles under ambient conditions," *Journal of Physical Chemistry C*, vol. 115, no. 14, pp. 6994–7001, Apr. 2011, doi: 10.1021/jp108797p.

[49] K. Sugikawa, S. Nagata, Y. Furukawa, K. Kokado, and K. Sada, "Stable and functional gold nanorod composites with a metal-organic framework crystalline shell," *Chemistry of Materials*, vol. 25, no. 13, pp. 2565–2570, Jul. 2013, doi: 10.1021/cm302735b.

[50] Q. L. Zhu, J. Li, and Q. Xu, "Immobilizing metal nanoparticles to metal-organic frameworks with size and location control for optimizing catalytic performance," *J Am Chem Soc*, vol. 135, no. 28, pp. 10210–10213, Jul. 2013, doi: 10.1021/ja403330m.

[51] Q. Yang *et al.*, "Regulating the spatial distribution of metal nanoparticles within metal-organic frameworks to enhance catalytic efficiency," *Nat Commun*, vol. 8, Feb. 2017, doi: 10.1038/ncomms14429.

[52] L. Chen, R. Luque, and Y. Li, "Encapsulation of metal nanostructures into metal-organic frameworks," *Dalton Transactions*, vol. 47, no. 11, pp. 3663–3668, 2018, doi: 10.1039/c8dt00092a.

[53] C. A. Celaya, A. G. el Hachimi, L. E. Sansores, and J. Muñiz, "Tailoring nanostructured materials based on γ-graphyne monolayers modified with Au heteroatoms for application in energy storage devices: A first principle study," *Appl Surf Sci*, vol. 598, Oct. 2022, doi: 10.1016/j.apsusc.2022.153771.

[54] A. Schneemann *et al.*, "Nanostructured Metal Hydrides for Hydrogen Storage," *Chemical Reviews*, vol. 118, no. 22. American Chemical Society, pp. 10775–10839, Nov. 28, 2018. doi: 10.1021/acs.chemrev.8b00313.

[55] V. Duc Chinh, G. Speranza, C. Migliaresi, N. van Chuc, V. Minh Tan, and N. T. Phuong, "Synthesis of Gold Nanoparticles Decorated with Multiwalled Carbon Nanotubes (Au-MWCNTs) via Cysteaminium Chloride Functionalization," *Sci Rep*, vol. 9, no. 1, Dec. 2019, doi: 10.1038/s41598-019-42055-7.

[56] B. I. Kharisov, O. v. Kharissova, U. Ortiz Méndez, and I. G. de La Fuente, "Decoration of Carbon Nanotubes With Metal Nanoparticles: Recent Trends," *Synthesis and Reactivity in Inorganic, Metal-Organic and Nano-Metal Chemistry*, vol. 46, no. 1, pp. 55–76, Jan. 2016, doi: 10.1080/15533174.2014.900635.

[57] S. Alim, J. Vejayan, M. M. Yusoff, and A. K. M. Kafi, "Recent uses of carbon nanotubes & gold nanoparticles in electrochemistry with application in biosensing: A review," *Biosensors and Bioelectronics*, vol. 121. Elsevier Ltd, pp. 125–136, Dec. 15, 2018. doi: 10.1016/j.bios.2018.08.051.

[58] C. B. Breslin, D. Branagan, and L. M. Garry, "Electrochemical detection of Cr(VI) with carbon nanotubes decorated with gold nanoparticles," *J Appl Electrochem*, vol. 49, no. 2, pp. 195–205, Feb. 2019, doi: 10.1007/s10800-018-1259-2.

[59] J. M. Soler *et al.*, "The SIESTA method for ab initio order-N materials simulation," 2002.

[60] J. P. Perdew, K. Burke, and M. Ernzerhof, "Generalized Gradient Approximation Made Simple," 1996.

[61] E. Artacho *et al.*, "SIESTA Steering Committee: Contributors to SIESTA," 2020. [Online]. Available: https://siesta-project.org

# Supplementary Information

## *Enhanced Hydrogen Storage in Gold-doped Carbon Nanotubes: A first-principles study*

Shima Rezaie [a], David M. J. Smeulders, [a,b] Azahara Luna-Triguero,*[a,b]

[a] *Energy Technology, Department of Mechanical Engineering, Eindhoven University of Technology, P.O. Box 513, 5600 MB Eindhoven, The Netherlands.*

[b] *Eindhoven Institute for Renewable Energy Systems (EIRES), Eindhoven University of Technology, PO Box 513, Eindhoven 5600 MB, The Netherlands*

## Table of Contents                Page

*Table S5. Glossary.*

| Symbol | Explanation | units |
|---|---|---|
| $E_b$ | Binding energy | eV |
| $E_f$ | Formation energy | eV |
| $E_F$ | Fermi energy | eV |
| $E_{tot(CNT)}$ | Total energy of carbon nanotube per unit cell | eV |
| $E_{tot(nAu)}$ | Total energy of *n* gold atoms | eV |
| $E_{tot(nAu-CNT)}$ | Total energy of carbon nanotube doped with *n* gold atom per unit cell | eV |
| $E_{tot(1Au-CNT)}$ | Total energy of *Au-doped CNT* per unit cell | eV |
| $E_{tot(C(graph))}$ | Total energy of the most stable structure of carbon (graphene) per unit cell | eV |
| $E_{tot(Au(bulk))}$ | Total energy of the most stable structure of gold (bulk) per unit cell | eV |
| $E_{tot(iH_2)}$ | Total energy of *i* hydrogen molecules | eV |
| $E_{tot((nAu-CNT)+iH_2)}$ | Total energy of the i adsorbed hydrogen molecules on the surface of carbon nanotube doped with n gold atom per unit cell | eV |
| $E_{ads}$ | Adsorption energy | eV |
| $q_t$ | Charge transfer | e |
| $q_{(ads-H_2)}$ | Total charge of the adsorbed hydrogen molecule | e |
| $q_{(iso-H_2)}$ | Total charge of the isolated hydrogen molecule | e |
| $T_d$ | Desorption temperature | K |
| wt % | Gravimetric storage capacity | - |
| $W_{H_2}$ | Molecular weight of hydrogen molecule | g/mol |
| $W_{(nAu-CNT)}$ | Molecular weight of nAu-doped CNT | g/mol |
| $M_V$ | Molecular weight of atoms per unit cell | g/mol |
| $n_v$ | Volumetric storage capacity | g/l |
| $\rho_{crystal}$ | Density of crystal per unit cell | g/l |
| $n$ | Number of gold atoms | - |
| $m$ | Number of carbon atoms | - |
| $i$ | Number of hydrogen molecules | - |
| Z | Number of atoms per unit cell | - |
| $k_B$ | Boltzmann constant | $1.38 \times 10^{-23}\ J/K$ |
| R | Gas constant | 8.314 J/(mol K) |
| $N_A$ | Avogadro constant | $6.023 \times 10^{23}$ $mol^{-1}$ |
| $p$ | Atmospheric pressure | 1 atm |
| $\Delta S$ | Change in hydrogen entropy from gas to liquid phase | 75.44 J/(mol.K) |
| $V_C$ | Volume of the unit cell | $cm^3$ |

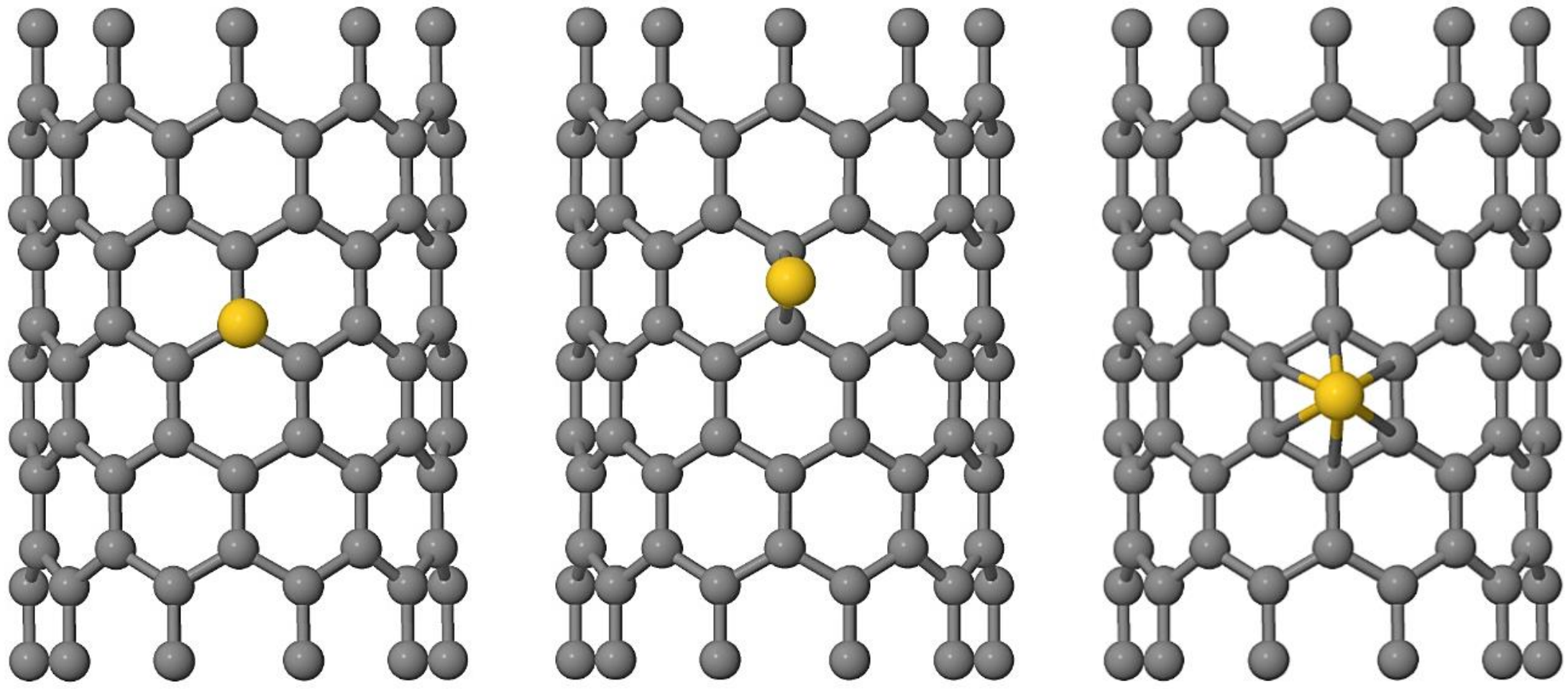

***Figure S4.*** *Different positions of gold on the surface of carbon nanotube.*

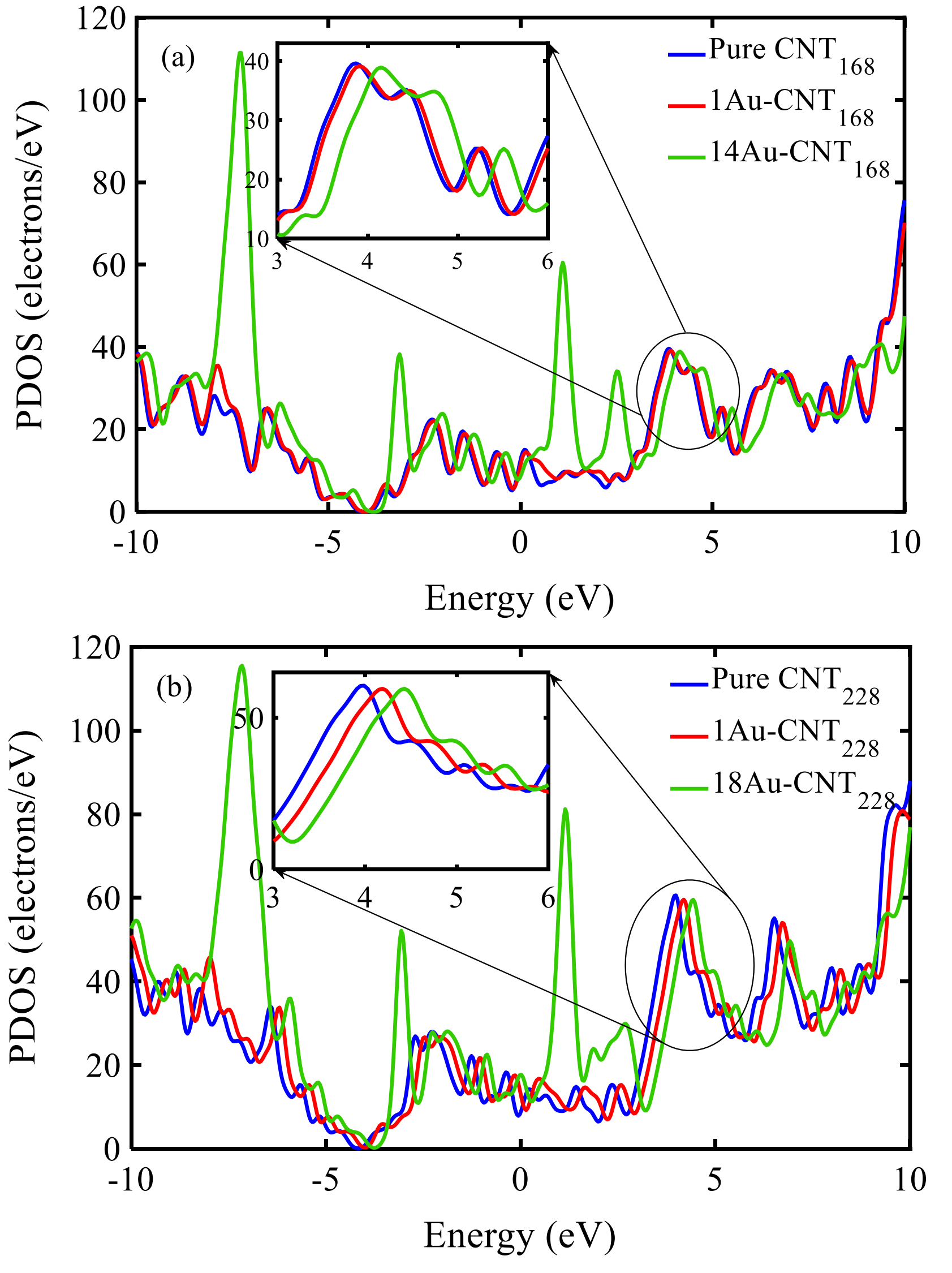

***Figure S5.*** *Total PDOS of a) $CNT_{168}$ and b) $CNT_{228}$ decorated with gold atoms. The surface is saturated for n=14 and 18 gold impurities.*

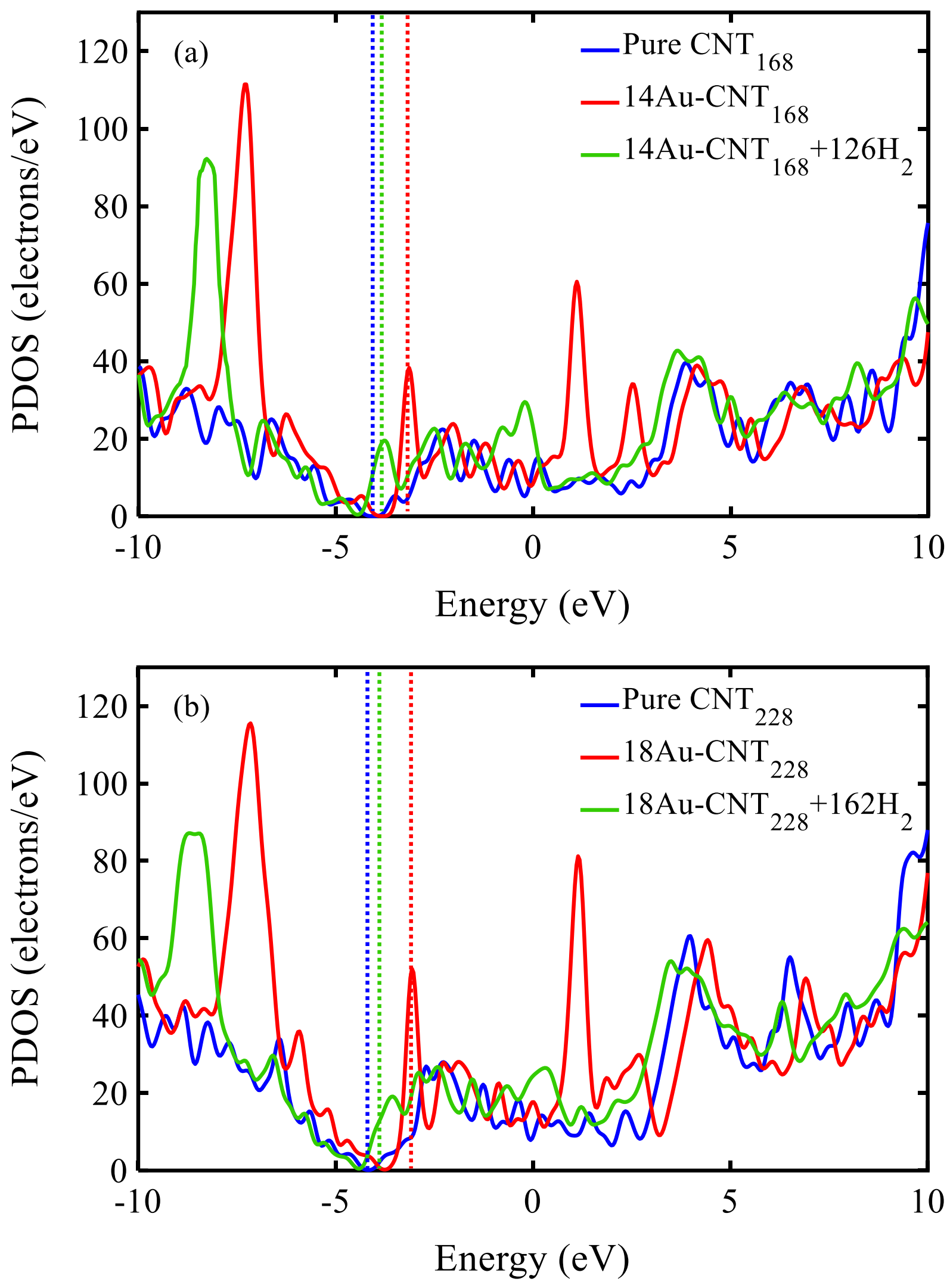

***Figure S6.*** *PDOS of pristine and saturated a) $CNT_{168}$ and b) $CNT_{228}$ with gold atoms, before and after the adsorption of hydrogen gases. Dashed lines indicate the Fermi Energy ($E_F$) per configuration.*

***Table S6.*** *Binding energy (eV) of hydrogen adsorption on carbon nanotubes with different diameters.*

| | **Vertical/Outside** | **Vertical/Inside** | **Horizontal/Outside** | **Horizontal/Inside** |
|---|---|---|---|---|
| **$CNT_{120}$** | 0.094 | 0.159 | 0.070 | 0.151 |
| **$CNT_{168}$** | 0.114 | 0.128 | 0.088 | 0.120 |
| **$CNT_{228}$** | 0.111 | 0.121 | 0.086 | 0.102 |

***Table S7.*** *Binding energy (eV) of hydrogen adsorption on 1Au-CNTs with different diameters.*

| | **1** | **2** | **3** | **4** | **5** | **6** | **7** | **8** | **9** | **10** |
|---|---|---|---|---|---|---|---|---|---|---|
| **$CNT_{120}$** | 0.217 | 0.223 | 0.211 | 0.219 | 0.214 | 0.213 | 0.153 | 0.138 | 0.127 | 0.118 |
| **$CNT_{168}$** | 0.235 | 0.221 | 0.231 | 0.217 | 0.203 | 0.192 | 0.187 | 0.172 | 0.164 | 0.142 |
| **$CNT_{228}$** | 0.238 | 0.224 | 0.220 | 0.219 | 0.201 | 0.195 | 0.183 | 0.173 | 0.164 | 0.151 |

***Table S8.*** *Desorption temperatures (K) of 1Au-CNTs with different diameters.*

| | **1** | **2** | **3** | **4** | **5** | **6** | **7** | **8** | **9** | **10** |
|---|---|---|---|---|---|---|---|---|---|---|
| **$CNT_{120}$** | 278 | 285 | 270 | 280 | 274 | 273 | 197 | 169 | 155 | 128 |
| **$CNT_{168}$** | 229 | 285 | 296 | 278 | 260 | 247 | 240 | 220 | 211 | 187 |
| **$CNT_{228}$** | 305 | 287 | 282 | 280 | 258 | 250 | 235 | 222 | 210 | 190 |